# Scaling of submicrometric Turing patterns in concentrated growing systems

Gabriel Morgado[1,2], Bogdan Nowakowski[1], and Annie Lemarchand[2]

[1]Institute of Physical Chemistry, Polish Academy of Sciences, Kasprzaka 44/52, 01-224 Warsaw, Poland
[2]Sorbonne Université, CNRS UMR 7600, Laboratoire de Physique Théorique de la Matière Condensée, 4 place Jussieu, case courrier 121, 75252 Paris CEDEX 05, France

Corresponding author: Annie Lemarchand, E-mail: anle@lptmc.jussieu.fr, Phone: 33 1 44 27 44 55

**Abstract**

The wavelength of a periodic spatial structure of Turing type is an intrinsic property of the considered reaction-diffusion dynamics and we address the question of its control at the microscopic scale for given dynamical parameters. The direct simulation Monte Carlo method, initially introduced to simulate particle dynamics in rarefied gases, is adapted to the simulation of concentrated solutions. We perform simulations of a submicrometric Turing pattern with appropriate boundary conditions and show that taking into account the role of the solvent in the chemical mechanism allows us to control the wavelength of the structure. Typically, doubling the concentration of the solution leads to decreasing the wavelength by two. The results could be used to design materials with controled submicrometric properties in chemical engineering. They could also be considered as a possible interpretation of proportion preservation of embryos in morphogenesis.

# 1   Introduction

The rational design of complex materials with predefined properties by controlling self-organization in far-from-equilibrium conditions is the promise of mesoscale chemical engineering [1, 2, 3]. Biological structures provide fascinating examples of organization and the concepts introduced to model them offer a good guide in an engineering context [4]. In particular, the ability to control the association and denaturation kinetics of nucleic acids has been successfully harnessed to link biological organization and material design [5, 6, 7, 8]. The model introduced by Turing [9] to account for periodic spatial structures in living organisms offers another example of mechanism inspired by biology that deserves to be re-examined with the goal of creating functional materials by biomimicry [10, 11, 12]. Turing proposed to model morphogenesis in the framework of far-from-equilibrium reaction-diffusion systems. Remarkably, Turing introduced a small number of processes based on a microscopic interpretation. A so-called activator is produced by an autocatalytic reaction while an inhibitor, associated with a larger diffusion coefficient, is consumed. An inhomogeneous perturbation is capable of destabilizing the homogeneous steady state in favor of a periodic spatial structure, whose wavelength is determined by the rate constants and the diffusion coefficients. There has been a recent revival of Turing's idea [13, 14, 15, 16, 4, 2, 10] and experimental evidence of Turing mechanism during embryogenesis has been provided [17, 18, 19]. However, the lack of scaling properties and versatility of Turing patterns is the main objection in biology as well as chemical engineering. A model of morphogenesis should reflect the adaptation of the wavelength to the size of the embryo. Similarly, the chemical engineer expects to adjust the wavelength of the spatial structure by monitoring an easily controlled parameter while considering the same chemical species.

The problem of wavelength scaling in a Turing pattern has been addressed for many years [20, 21, 22, 23] and remains topical [24, 25, 26, 27, 28, 29, 16]. Various scaling mechanisms have been proposed, mainly on a macroscopic scale. Partial differential equations with concentration-dependent diffusion terms [23, 27, 28] or involving an additional chemical species whose concentration is supposed to depend on system size [20, 21, 24, 25, 26, 29, 30] have been considered. Nevertheless, the control of a periodic spatial structure at a submicrometric scale in chemical engineering requires the design of a mechanism based on elementary processes, compatible with the simulation of particle dynamics. We recently proposed to place the problem of wavelength adaptation [31] in the context of molecular crowding [32, 33, 34, 35, 36, 37]. Frequently, solvent involvement in the chemical mechanism cannot be ignored in a concentrated solution [38]. We proposed a mechanism relying on microscopic processes in which the solvent is considered as a reagent in itself. In particular, the model does not introduce any ad hoc spatially-dependent variable. We studied the effect of the deviation from a dilute solution on the wavelength of a Turing pattern using partial differential equations. The perturbation of diffusion induced by crowding has been shown to have little effect on the wavelength of the structure [31].

In this paper, we wish to address the question of wavelength adaptation in a submicroscopic Turing pattern in the framework of chemical engineering. It implies designing an algorithm of particle dynamics simulations in a concentrated system. Turing model requires that the activator and the inhibitor have sufficiently different diffusion coefficients. To this goal, we adapt the procedure developed in a dilute system with three species of different diameters [11]. The results obtained in the high dilution limit reveals that, in a small system, the wavelength may be influenced by the boundary conditions. In particular, particle dynamics simulations of a given system with zero-flux boundary conditions and a length slightly smaller than two wavelengths lead to the selection of either one-and-a-half-wavelength or two-wavelength Turing pattern [11]. To avoid this kind of boundary effects, we take advantage of the specific conditions chosen to model the formation of the spine in a vertebrate embryo in a macroscopic description [39, 40, 41]. In order to reproduce the growth and spatial organization of the embryo, we started from a step function between two steady states and generated a propagating wavefront in a growing system. Turing pattern was developing between a fixed boundary condition at the rostral end and a moving front, which does not impose constraints on the wavelength of the structure. Hence, modelling of somitogenesis suggested appropriate boundary conditions for a problem of chemical engineering.

The paper is organized as follows. In section 2, we present a reaction-diffusion model with an explicit effect of the solvent that fades in the high dilution limit. The particle dynamics simulation method is presented in section 3. The results are given in section 4. Section 5 is devoted to conclusion.

## 2 Model

We consider a reaction scheme inspired by the Schnakenberg [42] and the Gray-Scott model [43]

$$A + S \xrightarrow{k_1^S} 2S \tag{1}$$

$$2A + B \xrightarrow{k_2} 3A \tag{2}$$

$$B + S \xrightarrow{k_3^S} 2S \tag{3}$$

$$\begin{cases} R \xrightarrow{k_{-3}} B \\ S \longrightarrow R \end{cases} \tag{4}$$

in which the role of the solvent S has been made explicit in Eqs. (1,3) and where $k_1^S$, $k_2$, $k_3^S$ and $k_{-3}$ are rate constants. The model involves two chemical species, the activator A and the inhibitor B, playing the role of two morphogens in somitogenesis. The reaction given in Eq. (2) consumes the inhibitor and autocatalytically produces the activator. The system is in contact with a reservoir R of species B which maintains the system far from equilibrium. The reaction given in Eq. (4) consists of two steps, an injection of species B at constant rate $k_{-3}$ and the simultaneous removal of the solvent S. The total concentration

$$C = A(x,t) + B(x,t) + S(x,t) \tag{5}$$

where $A(x,t)$, $B(x,t)$, and $S(x,t)$ are the concentrations of species A, B, and S, respectively, is constant. In the following, the spatial and temporal dependence of the concentrations is implicit. In the framework of a macroscopic approach, the system obeys the following reaction-diffusion equations:

$$\frac{\partial A}{\partial t} = -k_1 A \left[1 - \frac{A+B}{C}\right] + k_2 A^2 B + D_A \frac{\partial^2 A}{\partial x^2} \tag{6}$$

$$\frac{\partial B}{\partial t} = k_{-3} - k_3 B \left[1 - \frac{A+B}{C}\right] - k_2 A^2 B + D_B \frac{\partial^2 B}{\partial x^2} \tag{7}$$

where Eq. (5) has been used to eliminate $S$ and where the effective rate constants $k_1 = k_1^S C$ and $k_3 = k_3^S C$ have been introduced. The parameters $D_A$ and $D_B$ are the diffusion coefficients associated with species A and B, respectively. In the limit of a large amount of solvent S,

$$\frac{A+B}{C} \ll 1 \tag{8}$$

Eqs. (6) and (7) become

$$\frac{\partial A}{\partial t} = -k_1 A + k_2 A^2 B + D_A \frac{\partial^2 A}{\partial x^2} \tag{9}$$

$$\frac{\partial B}{\partial t} = k_{-3} - k_3 B - k_2 A^2 B + D_B \frac{\partial^2 B}{\partial x^2} \tag{10}$$

These reaction-diffusion equations can be associated with the chemical scheme that provides the benchmark in the high dilution limit:

$$A \xrightarrow{k_1} R_1 \tag{11}$$

$$2A + B \xrightarrow{k_2} 3A \tag{12}$$

$$B \underset{k'_{-3}}{\overset{k_3}{\rightleftharpoons}} R_2 \tag{13}$$

where $R_1$ and $R_2$ are reservoirs and $k'_{-3} = \frac{k_{-3}}{R_2}$.

For appropriate parameter values, Eqs (6,7) admit a stable homogeneous steady state

$$A^0 = 0 \tag{14}$$

$$B^0 = \frac{C}{2}\left(1 - \sqrt{1 - 4\frac{k_{-3}}{k_3 C}}\right) \tag{15}$$

and a steady state $(A^T, B^T)$ given in the Appendix, evolving into a Turing pattern in the presence of inhomogeneous perturbations. The two steady states depend on the total concentration $C$ and the rate constant ratios $k_1/k_2$, $k_3/k_2$, and $k_{-3}/k_2$.

Our aim is to characterize the impact of the deviation from the dilution limit

$$\delta \equiv \frac{A^0 + B^0}{C} = \frac{1}{2}\left(1 - \sqrt{1 - 4\frac{k_{-3}}{k_3 C}}\right) \tag{16}$$

on the wavelength of the pattern. In a concentrated system, $\delta$ does not vanish and the system is described by the reactions given in Eqs. (1-4). In the high dilution limit, $\delta$ vanishes and the mechanism is given by Eqs. (11-13).

We perform a linear stability analysis of Eqs. (6,7) around the state $(A^T, B^T)$. To this goal, the Fourier transforms $A_q(t) = \int_{-\infty}^{\infty} A(x,t)e^{-iqx}dx$ and $B_q(t) = \int_{-\infty}^{\infty} B(x,t)e^{-iqx}dx$ are introduced, where $q$ is the Fourier mode. In the Fourier space, the linear stability operator $M$ is given by:

$$M = \begin{pmatrix} k_2\alpha - D_A q^2 & k_2 M_{12} \\ k_2 M_{21} & k_2\beta - dD_A q^2 \end{pmatrix} \tag{17}$$

with

$$d = D_B/D_A \tag{18}$$

$$\alpha = \frac{k_1}{k_2}\left(1 - \frac{B^T}{C}\right) \tag{19}$$

$$\beta = -\frac{k_{-3}}{k_2 B^T} + \frac{k_3 B^T}{k_2 C} \tag{20}$$

$$M_{12} = \frac{k_1 A^T}{k_2 C} + A^{T2} \tag{21}$$

$$M_{21} = \frac{k_3 B^T}{k_2 C} - 2\frac{k_1}{k_2}\left(1 - \frac{A^T + B^T}{C}\right) \tag{22}$$

where the steady states $(A^0, B^0)$ and $(A^T, B^T)$ are given in Eqs. (14,15) and Eqs. (58,59) of the Appendix, respectively. For appropriately chosen parameter values, the largest eigenvalue of the operator $M$

$$\mu_+ = \frac{1}{2}\left(k_2(\alpha + \beta) - D_A(1 + d)q^2 + \sqrt{[k_2(\alpha - \beta) - D_A(1 - d)q^2]^2 + 4k_2^2 M_{12} M_{21}}\right) \tag{23}$$

is real and positive. Equation (16), relating $\delta$ and $C$, is used to introduce the dependence on the deviation from the dilution limit in the expression of the eigenvalue. The steady state $(A^T, B^T)$ is then unstable with respect to inhomogeneous perturbations and the mode $q_{\max}$, which maximizes the eigenvalue $\mu_+$, is the most unstable Fourier mode:

$$q_{\max} = \sqrt{\frac{k_2}{D_A}} \sqrt{\frac{\beta - \alpha}{d-1} + \frac{d+1}{d-1}\sqrt{\frac{-M_{12}M_{21}}{d}}} \qquad (24)$$

The wavelength of the periodic structure is then given by:

$$\lambda = \frac{2\pi}{q_{\max}} \qquad (25)$$

In Eqs. (6,7), time can be scaled by $1/k_2$ and space by $\sqrt{D_A/k_2}$. Figure 1 shows the variation of the scaled eigenvalue $\mu_+/k_2$ with respect to the square of the scaled Fourier mode $D_A q^2/k_2$ for two values of the deviation $\delta$ from the dilution limit. Interestingly, the most unstable mode $q_{\max}$ depends on $\delta$. As the solution becomes more concentrated, $q_{\max}$ increases, i.e., the wavelength $\lambda$ decreases. Moreover, the maximum eigenvalue $\mu_{\max} = \mu_+(q_{\max}^2)$ decreases revealing that the deviation from the dilution limit tends to destabilize Turing pattern.

The goal of the paper is to reexamine these properties at the microscopic scale using particle dynamics simulations. Following the results given in Fig. 1, the effect of the deviation from dilution limit on a Turing pattern will be explored in the range $0 \leq \delta \leq 0.05$.

## 3  Particle dynamics simulations

We use the Direct Simulation Monte Carlo method (DSMC), introduced by Bird [44, 45] to simulate the dynamics of a dilute gas. The method relies on a kinetic Monte Carlo algorithm and consists of a direct simulation of the Boltzmann equation including fluctuations. We already adapted the method to the simulation of hard spheres with different diameters in order to reproduce sufficiently different diffusion coefficients for the activator and the inhibitor [11]. Obtaining different binary mutual diffusion coefficients requires a third type of encounter, whose role was played by the reservoir. Here, the reservoir plays also the role of the solvent involved in the chemical scheme.

Particles are hard spheres of mass $m = 1$ with continuous positions and velocities. The initial velocity of the particles is sampled from a Maxwellian distribution with $k_B T = 1$.

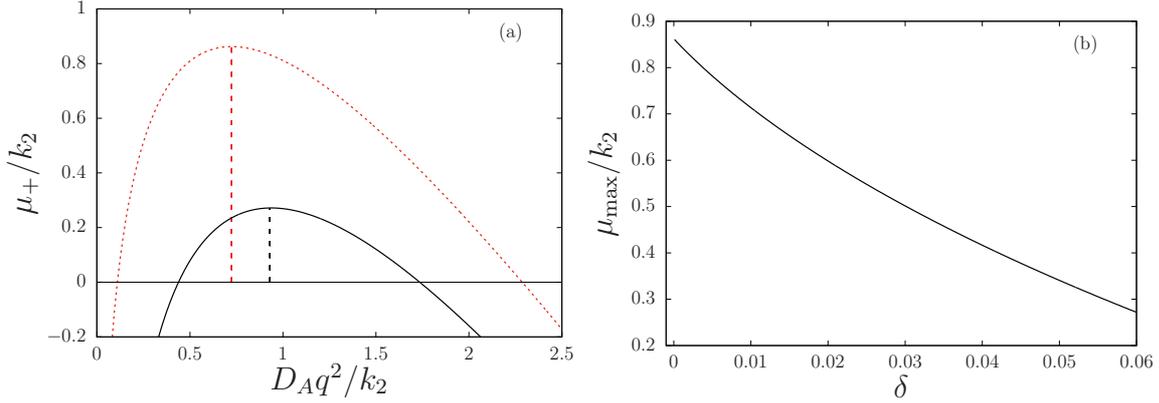

Figure 1: (a) Scaled positive eigenvalue $\mu_+/k_2$ of the linear stability operator given in Eq. (23) versus square of the scaled Fourier mode $D_A q^2/k_2$. Red dotted line: dispersion relation in the high dilution limit ($\delta = 0$). Black solid line: dispersion relation for $\delta = 0.05$. The square of the most unstable modes $q_{\max}^2$ are indicated by vertical dashed lines. (b) Maximum value of the scaled positive eigenvalue $\mu_{\max}/k_2$ versus the deviation $\delta$ from the dilution limit. The parameters take the following values: $\frac{k_1}{k_2} = 2.9 \times 10^4$, $\frac{k_3}{k_2} = 2.2 \times 10^4$, $\frac{k_{-3}}{k_2} = 8.8 \times 10^6$, $d = 10$.

During a time step, particle positions are updated according to their velocities. Updating of positions is performed along the $x$-axis whereas velocities are treated in a three-dimensional space. The treatment of the collisions requires space discretization. Only the particles belonging to the same spatial cell are susceptible to collide. Pairs of colliding particles are randomly chosen in a cell, according to the probability of collision. The latter is proportional to the relative velocity of the colliding pair, in agreement with the collision integral of the Boltzmann equation. Collisions are supposed to be elastic and the post-collision relative velocity is randomly chosen according to isotropic scattering.

We recall the procedure followed to obtain sufficiently different diffusion coefficients for species A and B in a ternary mixture of A, B, and S [11]. If the collisions A-B can be neglected with respect to the collisions A-S and B-S, i.e. if

$$SD_{AB} \gg AD_{BS} + BD_{AS}, \qquad S \gg B, \qquad S \gg A \qquad (26)$$

where $D_{XY}$ is the mutual diffusion coefficient in a binary mixture of X and Y, then the diffusion coefficients in the ternary mixture at local equilibrium obey [11, 46]

$$D_A \simeq D_{AS} = \frac{3}{8(A+S)(r_A+r_S)^2}\sqrt{\frac{k_BT}{\pi m}} \qquad (27)$$

$$D_B \simeq D_{BS} = \frac{3}{8(B+S)(r_B+r_S)^2}\sqrt{\frac{k_BT}{\pi m}} \qquad (28)$$

where $m$ is particle mass and $r_X$ is the radius of particles X=A, B, S. The choice

$$r_A = 2.2 r_S \qquad (29)$$

$$r_B = \frac{r_A + (1-\sqrt{d'})r_S}{\sqrt{d'}} \qquad (30)$$

where $d' = d\frac{C-A^T}{C-B^T}$, obeys the conditions given in Eq. (26) for a ratio $d = 10$ of the diffusion coefficients [11]. In the following, the radius of the solvent particles is set to $r_S = 0.5$. According to Eq. (30), the radius $r_B$ slightly varies with the chosen total concentration $C$, i.e. with the deviation $\delta$ from the dilution limit.

During a collision, a chemical reaction may occur according to the mechanisms described in Eqs. (1-4) for $\delta \neq 0$ and Eqs. (11-13) for $\delta = 0$. A collision between appropriate species is reactive with a probability proportional to the corresponding rate constant imposed by a steric factor. In order to save computation time, we only perform the collisions between species susceptible to react: we omit A-A, B-B, and S-S collisions that would have no effect on the chemical evolution of the system. The two chemical steps given in Eqs. (1,3) are standard binary reactions. For example, a collision A-S leads to the change of chemical nature of the A particle into a S particle at a frequency proportional to $k_1$. The exchanges with the reservoir R given in Eq. (4) are treated as follows: in each cell, randomly chosen S particles are turned into B particles at the constant rate $k_{-3}$, which simulates both the creation of a particle B and the simultaneous removal of a particle S at constant rate. We follow a well-accepted procedure [47] to treat the ternary reaction given in Eq. (2). The reaction is divided into two steps, the rate-limiting binary step, $A + B \xrightarrow{k_2} AB$, and the instantaneous reaction of the complex AB, $AB + A \longrightarrow 3\,A$. The formation of the complex AB is treated as a standard reactive binary collision with a condition on the relative velocity of the colliding pair. Considering that each cell is homogeneous, we evaluate the probability that the closest particle to the complex AB is of A type at $A/C$. Hence, the B particle is turned into an A particle with a probability equal to $A/C$. Consequently, the rate constant of the ternary reaction is given by:

$$k_2 = \frac{4(r_A + r_B)^2}{C}\sqrt{\frac{\pi k_B T}{m}} \qquad (31)$$

It is worth noting that the total concentration $C$ is related to the deviation $\delta$ from the dilution limit according to Eq. (16). Hence, the rate constant $k_2$ is affected by the dilution of the system. As already mentioned, the steady states $(A^0, B^0)$ and $(A^T, B^T)$ only depend on the ratios $k_1/k_2$, $k_3/k_2$, and $k_{-3}/k_2$. Consequently, we choose to assign constant values to these ratios, so that the steady states deduced from the simulations are expected to be identical to the steady states of the macroscopic description for all the studied values of $\delta$. In all the simulations, the rate constant ratios and the diffusion coefficient ratio are set to:

$$\frac{k_1}{k_2} = 2.9 \times 10^4, \qquad \frac{k_3}{k_2} = 2.2 \times 10^4, \qquad \frac{k_{-3}}{k_2} = 8.8 \times 10^6, \qquad d = 10 \qquad (32)$$

In the concentrated system of interest, the smallest mean free path,

$$\ell = \frac{1}{\sqrt{2}C\pi(r_A + r_S)^2} \qquad (33)$$

is associated with the collisions A-S. In dilute gases, cell length $\Delta x$ is chosen smaller than the mean free path. Here, this condition would lead to prohibitive computation times. In order to reach a sufficient spatial resolution, we choose $\Delta x = \frac{\lambda}{50}$ where the wavelength $\lambda$ of Turing pattern is evaluated according to Eq. (25). The variation of the concentration between two cells is sufficiently small to legitimate the choice of a cell length larger than the mean free path. Following the DSMC method [44, 45], we impose the time step $\Delta t = \frac{\tau}{5}$ where $\tau = \frac{\ell}{\bar{v}}$ is the mean free time and $\bar{v} = 2\sqrt{\frac{2k_B T}{\pi m}}$ is the mean speed. The system size is set at $L = 10\lambda$. For a given value of the deviation $\delta$ from the dilution limit, the total concentration $C$ is deduced from Eq. (16). The macroscopic initial condition is a step function between the steady state $(A^T, B^T)$ in the first 50 cells on the left and the steady state $(A^0, B^0)$ in the remaining cells. The concentration of the solvent S is deduced from the conservation relation given in Eq. (5). The initial numbers of particles A, B, and S in each simulation cell are the nearest integers to the corresponding real concentration values. Typically, for $\delta = 0.02$, we find the following initial numbers of particles in each cell prepared in the steady state $(A^T, B^T)$:

$$N^0_{A^T} = 194, \qquad N^0_{B^T} = 148, \qquad N^0_{S^T} = 20058 \qquad (34)$$

In these conditions, a Turing pattern develops behind a wave front propagating to the right. The propagation of the wave front is not disturbed by bulk nucleation. Indeed, the chemical mechanism given in Eqs. (1-4) involves a reservoir of species B but not of species A. Contrary to species B, species A is not injected into the system, which ensures the existence of a steady state with a vanishing number of particles A per cell. Hence, the system prepared in the steady state $(A^0, B^0)$ cannot produce A particles, neither due to reaction nor injection, implying that the system remains in the state $(A^0 = 0, B^0)$ even in the presence of large fluctuations.

Cells that are more than $50\Delta x$ to the right of the wavefront are not updated to save computation time. Zero-flux boundary conditions are imposed which results in extremum concentrations for A and B species on the left boundary when a Turing pattern has developed. The simulation is stopped as the front traveled $0.75L$.

In the next section, we examine how the wavelength of the Turing pattern deduced from the simulations varies with the deviation $\delta$ from the dilution limit. The results will be compared to the macroscopic predictions.

## 4 Results

Instantaneous spatial profiles associated with species A and B are given in Fig. 2 in the case of a concentrated system for a deviation $\delta = 0.05$ from the dilution limit. The numerical solution of the reaction-diffusion equations given in Eqs. (6,7) and the simulation results for the chemical mechanism given in Eqs. (1-4) are compared for the same parameter values. As expected, a Turing pattern has developed behind a wave front and the wavelengths obtained for the macroscopic description and the simulations agree well. However, the positions of the wave fronts corresponding to the deterministic approach and the simulations are different. The wave front associated with the simulations has a fluctuating position. Nevertheless, we obtain a larger propagation speed than the deterministic prediction for all the simulations.

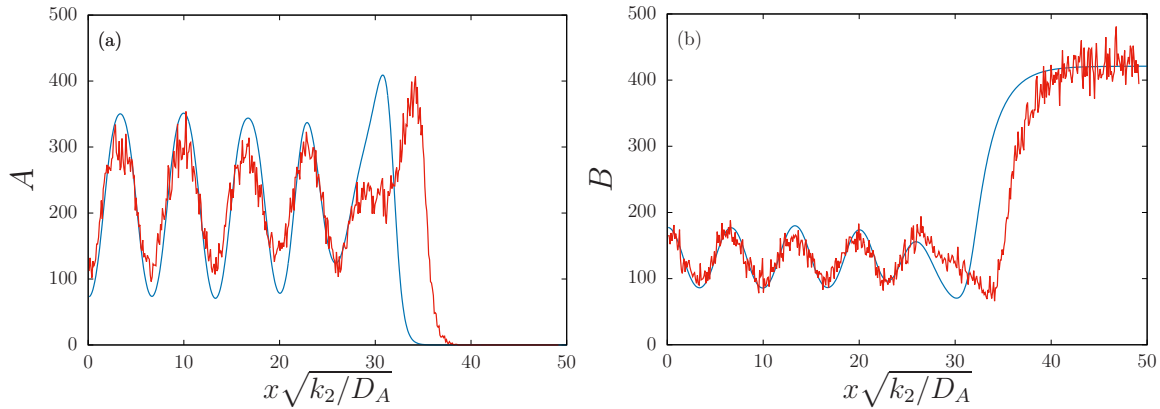

Figure 2: Concentration profiles deduced from the numerical solution of Eqs. (6,7) (blue) and number of particles per cell in simulations (red) versus scaled spatial coordinate $x\sqrt{\frac{k_2}{D_A}}$ for Eqs. (1-3), for a given deviation $\delta = 0.05$ from the dilution limit, at scaled time $k_2 t = 6986$. Snapshot of the profiles for species A (a) and B (b). The value of the other parameters is given in Sec. 3.

The macroscopic predictions of the steady state $(A^T, B^T)$ given in Eqs. (58,59) are compared to the spatially-averaged numbers $\langle N_{A^T} \rangle$ and $\langle N_{B^T} \rangle$ of particles A and B deduced from the simulations. The average is performed over four wavelengths. The results are displayed in Fig. 3 for different values of the deviation $\delta$ from the dilution limit. The simulation results are not displayed for $\delta = 0.01$ due to the prohibitive computation time necessary for the wave front to reach $7.5L$. The total number $N_C$ of particles per cell to be considered dramatically increases as $\delta \to 0$. For example, $N_C$ reaches 40400 for $\delta = 0.01$.

The mechanism involving the solvent as a reactive species given in Eqs. (1-4) has been used to perform the reactive collisions for $\delta \neq 0$. The mechanism associated with the high dilution limit given in Eqs. (11-13) has been used to treat the reactive collisions in the case $\delta = 0$, for which the total concentration $C$ can be arbitrary chosen since the solvent is chemically inert. We have checked that the mean numbers of particles $\langle N_{A^T} \rangle$ and $\langle N_{B^T} \rangle$ do not change in the explored range $7980 \leq N_C \leq 19950$ of total number $N_C$ of particles values. As a result, increasing $\delta$ widens the gap between the stationary values $A^T$ and $B^T$. The very good agreement between the macroscopic and simulation results can be considered as a test of the simulation procedure. However, the simulation results systematically slightly underestimate the gap between $\langle N_{A^T} \rangle$ and $\langle N_{B^T} \rangle$. The numerical cost prevents us from computing spatially-averaged numbers of particles for a sufficient number of wavelengths far enough from the wave front. As shown in Fig. 1, the front associated with species A has a negative gradient. Thus, $N_A$ decreases at the wavefront whereas $N_B$ increases. Consequently, the mean number $\langle N_{A^T} \rangle$ is slightly underestimated whereas $\langle N_{B^T} \rangle$ is slightly overestimated. The agreement between macroscopic and simulation results confirms that the simulation method correctly reproduces the ratios $k_1/k_2$, $k_3/k_2$, and $k_{-3}/k_2$ of the rate constants for all the considered values of the deviation $\delta$ from the dilution limit. This result *a posteriori* legitimates the choice of the cell length $\Delta x$ for which the standard requirements of DSMC have not been strictly obeyed.

The simulation method being validated, we now report on the main issue of the paper: the ability of tuning the wavelength of a Turing pattern by controlling the dilution of the system. The simulation model leads to non trivial dependences of the dynamical parameters on the deviation $\delta$ from the dilution limit. As shown in Eq. (27,31), the diffusion coefficient $D_A$ and the rate constant $k_2$ depends on the total concentration $C$ and, hence, on $\delta$. Nevertheless, the ratios $d$, $k_1/k_2$, $k_3/k_2$, and $k_{-3}/k_2$ remain constant in the simulations, regardless of the deviation $\delta$ from the dilution limit. According to Eqs. (24,25), the dimensionless quantity $\lambda\sqrt{\frac{k_2}{D_A}}$ associated with the macroscopic description depends on the constant ratios and $\delta$ in the same way as the corresponding quantity deduced from the simulations. Figure 4 presents the scaled wavelength $\lambda\sqrt{\frac{k_2}{D_A}}$ of the Turing pattern versus the deviation $\delta$ from the dilution limit. The macroscopic results are deduced from Eqs. (24,25) for $\lambda$ and Eqs. (27,31) for $D_A$ and $k_2$. Whereas the macroscopic prediction of the wavelength $\lambda$ computed according to Eqs. (24,25,27,31) diverges in the limit $\delta \to 0$, the scaled wavelength $\lambda\sqrt{\frac{k_2}{D_A}}$ tends to a finite value as shown in Fig. 4. The simulation results are obtained for the mechanism given in Eqs. (1-4) for a non vanishing value of the deviation $\delta$ from the dilution limit and for the mechanism given in Eqs. (11-13) for $\delta = 0$.

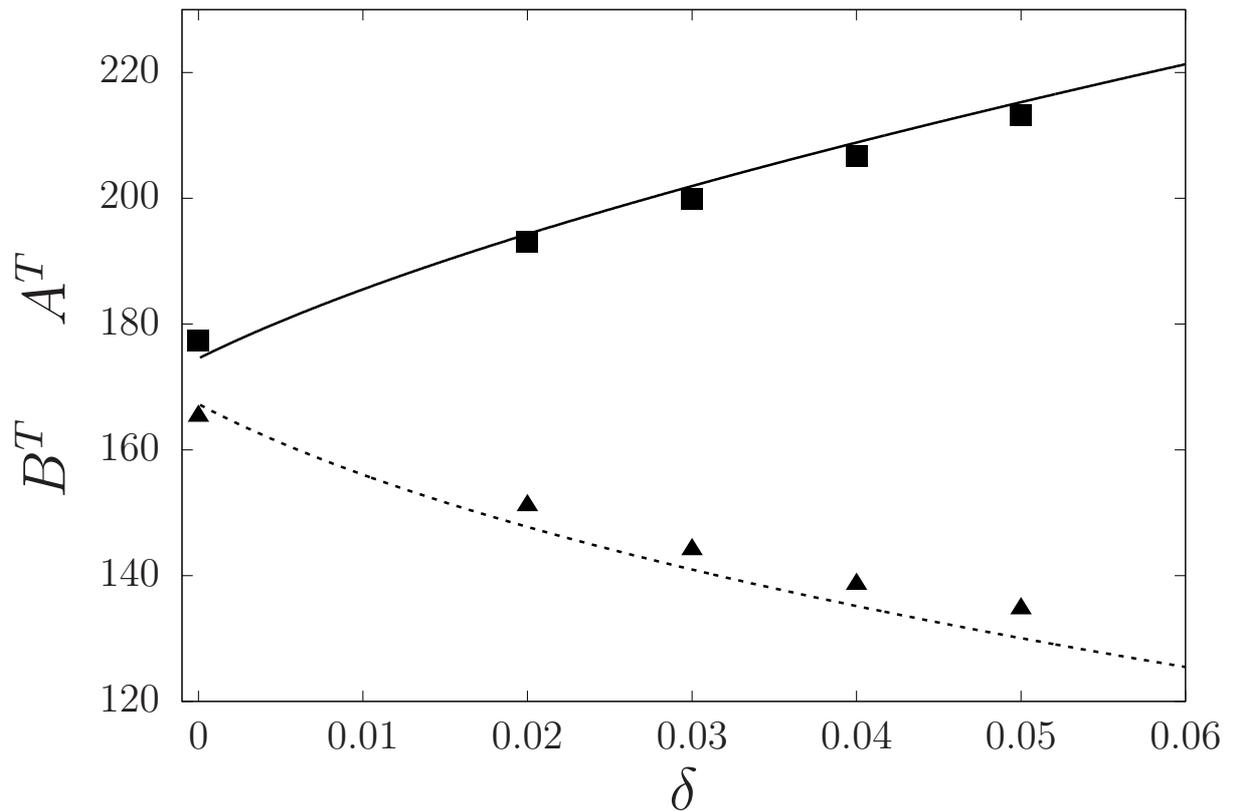

Figure 3: Macroscopic prediction of the steady state concentrations $A^T$ (solid line) given by Eq. (58) and $B^T$ (dotted line) deduced from Eq. (59) versus the deviation $\delta$ from the dilution limit. Squares (triangles, resp.) give the mean number of A (B, resp.) particles per cell in the region of the simulated Turing pattern. For $\delta \neq 0$, the mechanism given in Eqs. (1-4) is used whereas the mechanism given in Eqs. (11-13) is used for $\delta = 0$. The value of the other parameters is given in Sec. 3.

The same spatial averaging over four wavelengths has been applied to compute the scaled wavelength deduced from the simulations as for the determination of the steady states. The agreement between the macroscopic and simulation results about scaled wavelength versus $\delta$ is satisfactory. Nevertheless, the simulation results fluctuate more as $\delta$ increases. This result was expected due to the reduced stability of Turing structure observed in Fig. 1 when $\delta$ increases. Varying the dilution of the system enables adjusting the scaled wavelength of the Turing pattern. A 10% variation is reached when switching from $\delta = 0$ to $\delta = 0.05$ but only a 5% variation is obtained from $\delta = 0.02$ to $\delta = 0.05$.

However, the simulation results may be interpreted in a different way, in particular, in the perspective of chemical engineering. If we assume that it is possible to find an experimental chemical system obeying Eqs. (1-4) with the same dependences of the dynamical parameters as in the chosen simulation model, then the wavelength of the observed Turing pattern will follow the non scaled dependence on the deviation $\delta$ from the dilution limit. The results shown in Fig. 5 mimic the results that could be deduced from such an experimental system. As already mentioned, the wavelength $\lambda$ diverges as $\delta \to 0$. The variation between the wavelength values as $\delta$ increases from 0.02 to 0.05 reaches 67%. For such a large sensitivity of the wavelength to dilution, the fluctuations have a negligible effect and the simulation results perfectly agree with the macroscopic predictions. This excellent agreement between DSMC results and a macroscopic approach, which neglects the effect of the deviation from the dilution limit on diffusion, confirms that the main effect on the wavelength is induced by the perturbation of the reaction in a concentrated system. The simulation results have been given in arbitrary units. However, it is easy to evaluate the order of magnitude of the wavelength in nanometers. Typically, the simulations involve 350 particles A and B for 8500 particles in a cell in the case of a deviation $\delta = 0.05$ from the dilution limit. For small molecules A and B in water S, it amounts to total concentrations of the order of $C = 2.2$ mol/L, i.e. 1.4 particles per $nm^3$. The sum of the radii of the particles A and S is estimated at $r_A + r_S = 0.6$ nm. Using Eq. (33) for the mean free path $\ell$ and taking into account that we impose $\lambda \simeq 500\ell$, we find that the wavelength of the Turing pattern is in the order of 200 nanometers. The results of DSMC simulations including an explicit effect of the solvent in the chemical scheme show that varying dilution allows us to control the wavelength of a submicrometric spatial structure.

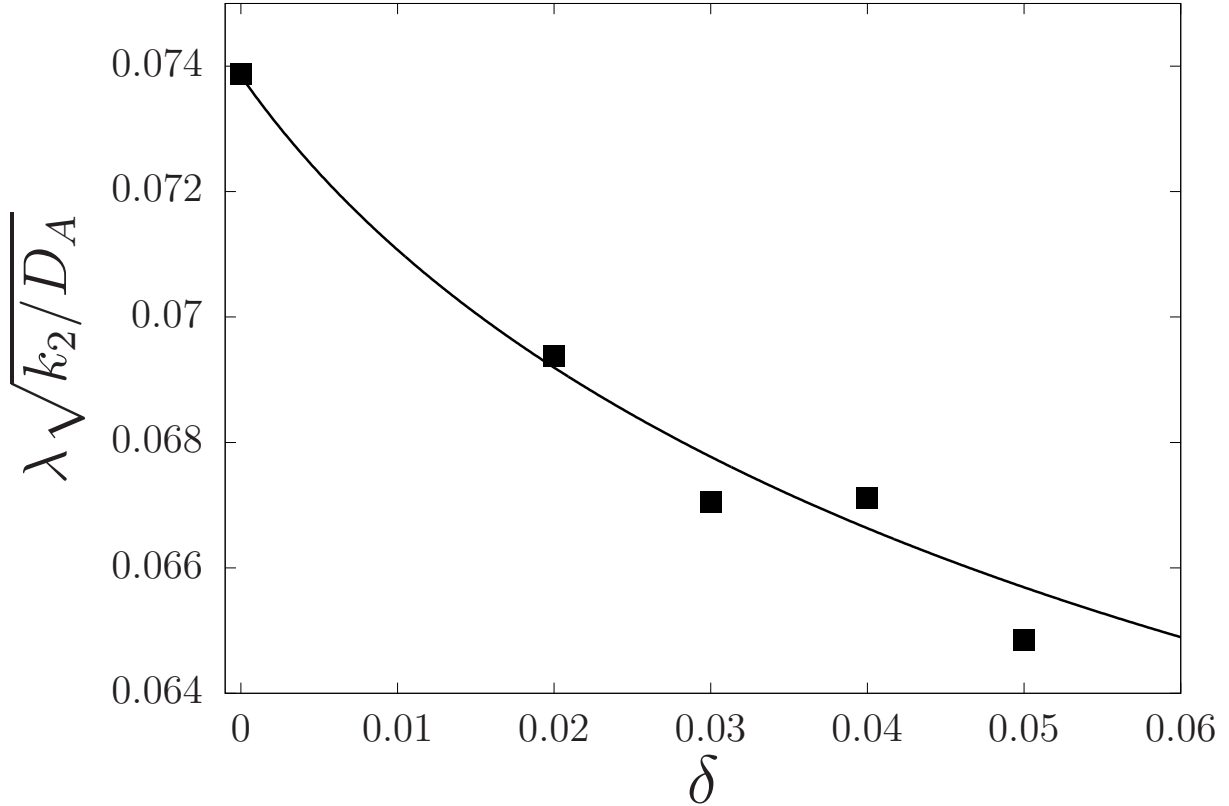

Figure 4: Macroscopic prediction of the scaled wavelength of the Turing pattern given by Eqs. (24,25,27,31) (solid line) and scaled wavelength deduced from the simulations (squares) versus deviation $\delta$ from the dilution limit. For $\delta \neq 0$, the mechanism given in Eqs. (1-4) is used whereas the mechanism given in Eqs. (11-13) is used for $\delta = 0$. The value of the other parameters is given in Sec. 3.

# 5 Conclusion

In this paper, we present a reaction-diffusion model based on elementary processes that enables the control of a spatially periodic pattern at the microscopic scale. The direct simulation Monte Carlo (DSMC) method has been extended to the simulation of concentrated solutions with the aim of generating Turing patterns with a tunable wavelength at a submicrometric scale. The wavelength of Turing pattern is imposed by dynamics and we show that the deviation from the dilution limit can be harnessed to adjust the wavelength to a selected value. Often, the role of the solvent as a reactive species cannot be ignored in concentrated solutions [38]. DSMC has been successfully used to show the possibility to monitor submicrometric Turing patterns by controlling the total concentration, provided that the chemical mechanism takes the solvent into account. In the high dilution limit, the considered reaction-diffusion equations converge to the equations associated with the chemical mechanism without explicit role of the solvent. We demonstrate that increasing the total concentration by a factor 2 is sufficient to obtain a wavelength reduction of the same factor.

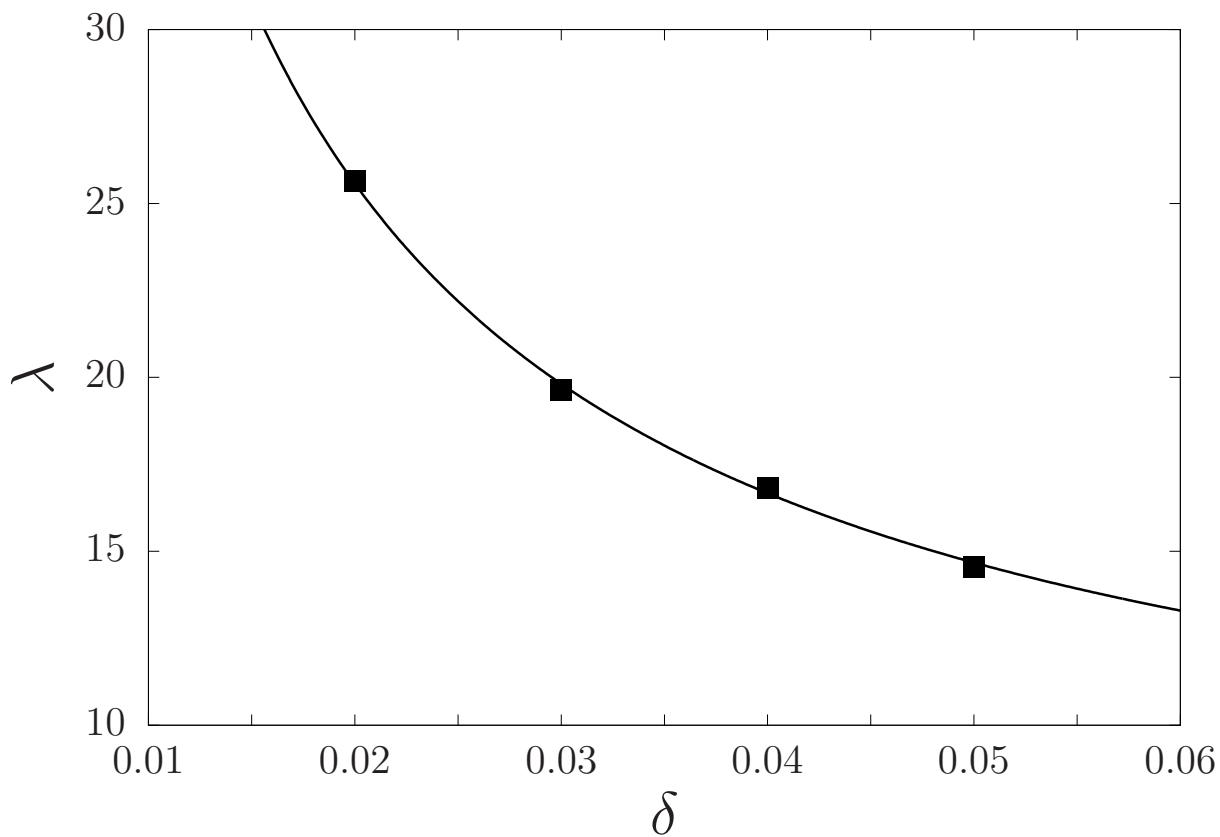

Figure 5: Same caption as in Fig. 4 without scaling of the wavelength.

The proposed scenario, involving a strengthening of molecular crowding in smaller embryos, could be considered to support the observed scaling of spatial structures in biology, for example, the adaptation of somite size to embryo size and, more generally, the preservation of proportions in morphogenesis [48]. The results give some hints to design a chemical scheme enabling the formation of a tailored Turing pattern in mesocale chemical engineering.

# Appendix

The Appendix is devoted to the derivation of the steady-state $(A^T, B^T)$:

$$0 = -\frac{k_1}{k_2}A^T\left[1 - \frac{A^T + B^T}{C}\right] + A^{T^2}B^T \tag{35}$$

$$0 = \frac{k_{-3}}{k_2} - \frac{k_3}{k_2}B^T\left[1 - \frac{A^T + B^T}{C}\right] - A^{T^2}B^T \tag{36}$$

associated with the mechanism given in Eqs. (1-4). It reads:

$$B^T = \frac{(k_1/k_2)(C - A^T)}{(k_1/k_2) + A^T C} \tag{37}$$

$$aA^{T^4} + bA^{T^3} + cA^{T^2} + dA^T + e = 0 \tag{38}$$

where

$$a = -\frac{k_1}{k_2}C \tag{39}$$

$$b = \frac{k_1}{k_2}\left(\frac{k_3}{k_2} - \frac{k_1}{k_2} + C^2\right) \tag{40}$$

$$c = -\frac{k_{-3}}{k_2}C^2 - \frac{k_1}{k_2}\left(2\frac{k_3}{k_2} - \frac{k_1}{k_2}C\right) \tag{41}$$

$$d = \frac{k_1}{k_2}\left(\frac{k_3}{k_2}C^2 - 2\frac{k_{-3}}{k_2}C\right) \tag{42}$$

$$e = -\frac{k_{-3}}{k_2}\frac{k_1}{k_2}\frac{k_1}{k_2} \tag{43}$$

Following Ferrari method [49], we introduce:

$$z = A^T + \frac{b}{4a} \tag{44}$$

$$p = \frac{-3b^2}{8a^2} + \frac{c}{a} \tag{45}$$

$$q = \frac{b^3}{8a^3} - \frac{bc}{2a^2} + \frac{d}{a} \tag{46}$$

$$r = -3\left(\frac{b}{4a}\right)^4 + \frac{b^2 c}{16a^3} - \frac{bd}{4a^2} + \frac{e}{a} \tag{47}$$

and write Eq. (38) in the form

$$z^4 + pz^2 + qz + r = 0 \tag{48}$$

It is sufficient to find a root of the third equation:

$$8y^3 - 4py^2 - 8ry + 4pr - q^2 = 0 \tag{49}$$

to find the roots of Eq. (48). Following Cardan method [49], we introduce

$$a' = 8 \; ; \; b' = -4p \; ; \; c' = -8r \; ; \; d' = 4rp - q^2$$

$$p' = -\frac{b'^2}{3a'^2} + \frac{c'}{a'} \; ; \; q' = \frac{b'}{27a'}\left(\frac{2b'^2}{a'^2} - \frac{9c'}{a'}\right) + \frac{d'}{a'}$$

and the discriminant of Eq. 49 reads:

$$\Delta_3 = -(4p'^3 + 27q'^2) \tag{50}$$

For the parameters given in Eq. (32), the discriminant $\Delta_3$ is positive and Eq. (49) has three real solutions, one of which obeys

$$y^0 = u + \bar{u} - \frac{b'}{3a'} \tag{51}$$

with $u = \left(\dfrac{-q' + i\cdot\sqrt{\dfrac{\Delta_3}{27}}}{2}\right)^{1/3}$. The quartic polynomial given in Eq. (48) can be factorized in two quadratic polynomials associated with the following discriminants:

$$\Delta_4^{(1)} = -2y^0 - p + \frac{2q}{\sqrt{2y^0 - p}} \tag{52}$$

$$\Delta_4^{(2)} = -2y^0 - p - \frac{2q}{\sqrt{2y^0 - p}} \tag{53}$$

The four solutions of Eq. 38 are given by:

$$A^{T(1)} = \frac{1}{2}\left(-\sqrt{2y^0} + \sqrt{\Delta_4^{(1)}}\right) - \frac{b}{4a} \tag{54}$$

$$A^{T(2)} = \frac{1}{2}\left(-\sqrt{2y^0} - \sqrt{\Delta_4^{(1)}}\right) - \frac{b}{4a} \tag{55}$$

$$A^{T(3)} = \frac{1}{2}\left(\sqrt{2y^0} + \sqrt{\Delta_4^{(2)}}\right) - \frac{b}{4a} \tag{56}$$

$$A^{T(4)} = \frac{1}{2}\left(\sqrt{2y^0} - \sqrt{\Delta_4^{(2)}}\right) - \frac{b}{4a} \tag{57}$$

The solution sought must converge toward the known expression of the steady state [11] in the limit of a diluted system with $C \gg A + B$. Only $A^{T(4)}$ obeys the previous requirement. We find:

$$A^T = \frac{1}{2}\left(\sqrt{2y^0} - \sqrt{\Delta_4^{(2)}}\right) - \frac{b}{4a} \tag{58}$$

$$B^T = \frac{(k_1/k_2)(C - A^T)}{(k_1/k_2) + A^T C} \tag{59}$$

Equations (58,59) are used to draw Fig. 3.

# Acknowledgements

This publication is part of a project that has received funding from the European Union's Horizon 2020 research and innovation programme under the Marie Skłodowska-Curie grant agreement No. 711859. Scientific work funded from the financial resources for science in the years 2017-2021 awarded for the implementation of an international co-financed project.